\documentclass[preprint,12pt]{aastex}
\usepackage{graphicx}

\shorttitle{GD-1 Orbit}
\shortauthors{Willett et al.}

\begin{document}

\title{An Orbit Fit for the Grillmair Dionatos Cold Stellar Stream}
\author{
Benjamin A. Willett,\altaffilmark{\ref{RPI}}
Heidi Jo Newberg,\altaffilmark{\ref{RPI}}
Haotong Zhang,\altaffilmark{\ref{NAO}}
Brian Yanny,\altaffilmark{\ref{fermi}}
Timothy C. Beers\altaffilmark{\ref{msu}}
}

\altaffiltext{1}{Dept. of Physics, Applied Physics, and Astronomy, Rensselaer Polytechnic Institute, Troy, NY 12180; willeb@rpi.edu\label{RPI}}
\altaffiltext{2}{National Astronomical Observatory, Beijing, China;\label{NAO}}
\altaffiltext{3}{Fermi National Accelerator Laboratory, Batavia, IL 60510; \label{fermi}}
\altaffiltext{4}{Dept. of Physics and Astronomy, CSCE: Center for the Study of Cosmic Evolution, and JINA: Joint Institute for Nuclear Astrophysics, Michigan State University, E. Lansing, MI  48824, USA; beers@pa.msu.edu \label{msu}}


\pagestyle{plain}

\begin{abstract}
	We use velocity and metallicity information from SDSS and SEGUE stellar spectroscopy 
to fit an orbit to the narrow $63^\circ$ stellar stream of Grillmair and Dionatos.
The stars in the stream have a retrograde orbit with eccentricity $e = 0.33$ (perigalacticon
of 14.4 kpc and apogalacticon of 28.7 kpc) and inclination approximately $i \sim 35^\circ$.  
In the region of the orbit which is detected, it has a distance of about 7 to 11 kpc 
from the  Sun.  Assuming a standard disk plus bulge and 
logarithmic halo potential for the Milky Way stars plus dark matter,
the stream stars are moving with a large space velocity of approximately $276~ \rm km~s^{-1}$ at
perigalacticon.  Using this stream alone, we are unable to determine 
if the dark matter halo is oblate or prolate.   The metallicity of the stream 
is [Fe/H] $= -2.1\pm0.1$.  Observed proper motions for individual stream members above the
main sequence turnoff are consistent with the derived orbit.
None of the known globular clusters in the Milky Way have positions, radial velocities, and
metallicities that are consistent with being the progenitor of the GD-1 stream.
\end{abstract}

\section{Introduction}

Data from tidal stream debris is a valuable resource for constraining Galactic structure.  
In the last decade, several streams, with both globular cluster and dwarf galaxy progenitors,
have been discovered in the Milky Way 
\citep{ibata95,leon00,ynetal00,vetal01,nyetal02,martinezdelgado01, odenkirchen01,majewski03,gj06,detal06,gd06,g06,betal07,Grill09}, 
as well as in nearby galaxies \citep{ietal01a,2005ApJ...622L.109F,chapman08,martinezdelgado08,mdetal08}. 
By examining the density, kinematics, distribution, and structure of 
various tidal streams surrounding the Milky Way, a clearer picture of how our halo 
was built can be developed \citep{bj05}.  In addition, stellar streams can be used as probes of the Galactic
gravitational potential, and thus constrains the shape of the dark matter halo
 \citep{ilitq01,h04,fellhauer}.

\citet{gd06}, hereafter GD, announced the detection of a $63^\circ$ cold stellar stream 
in the Galactic halo (the stream itself we refer to henceforth as GD-1, following GD), 
using stellar density counts extracted from the Sloan Digital Sky Survey (SDSS; York et al. 2000).  
This stream is extremely narrow, less than $0.25^\circ$ degrees in width, which is less than 50 pc 
at their measured distances of 7.3 to 9.1 kpc from the Sun.  GD therefore concluded that
the progenitor was a globular cluster, but the progenitor remains unidentified and
could be completely disrupted.

In this work, we utilize newly available Sloan Extension for Galactic Understanding and Exploration (SEGUE; see Yanny et al. 2009) spectroscopy of stars identified along 
the stream, that are available in the Sloan Digital Sky Survey (SDSS) Data Release 7 (DR7), to constrain 
the orbit of the stream and search for possible progenitors.

\section{Data Selection}

The SDSS imaging survey \citep{gunn98, gunn06} has made it possible to detect faint Milky Way halo
streams from the spatial distribution of stars because it provides very accurate multicolor photometry
for millions of Galactic stars.
Even with the well-calibrated SDSS photometry \citep{fukugita96,smith02,tucker06,hogg01,pier03}, GD
separated the faint structure from the background of Milky Way halo and disk stars
only with careful filtering and smoothing techniques \citep{odenkirchen01,matched}.
To characterize the stream in more detail and to compute an orbit, we reanalyze 
the imaging data, supplemented through DR7, and then add to it all newly 
available spectroscopic observations of GD-1 stream stars available in SDSS DR7.

\subsection{Imaging}

Using the analysis of GD as a guide, stars were selected from the
SDSS DR7 \citep{dr7} footprint from a rough color-magnitude box restricted 
to blue F turnoff and upper main sequence stars at distances of
7 to 20 kpc from the Sun (taking into account that the turnoff spans more than
a magnitude of absolute magnitudes): $0 < (g-r)_0 < 0.5, 18 < g_0 < 22$.  
Magnitudes with $0$ subscript indicate those which have been corrected for 
reddening using the \cite{sfd98} maps as implemented in SDSS (DR7).
All SDSS stars in the DR7 North Galactic cap footprint which are in this
color-magnitude box were selected and plotted in an ($\alpha,\delta$) 
stellar density diagram with pixels $0.5^\circ$ on a side.
Examination of this diagram by eye showed that GD-1 stream stood out weakly from the 
background, with enough contrast so that the location of several fiducial points 
along the stream in ($\alpha,\delta$) could be mapped.
Low-order polynomials were then fit to the positions of these points.
The lowest order best fit was of third order:
\begin{equation}
\delta =  -864.5161+13.22518 \alpha-0.06325544 \alpha^2+  0.0001009792 \alpha^3 \label{eqn1}.
\end{equation}
By comparison with likely identified spectroscopic stream members we later
verified that this polynomial matches the stream position within
about $0.6^\circ$ for $130^\circ < \alpha < 175^\circ$, and within about $1.0^\circ$ for
$120^\circ < \alpha < 130^\circ$ and $175^\circ < \alpha < 220^\circ$.

Next, 147,537 stars within $\pm 0.5^\circ$ of this polynomial fit (a conservative width is used to increase the signal to noise, it is not necessary
to identify every star) were selected
from $135^\circ < \alpha < 200^\circ$. They represent the `on-stream' data set.
For a control sample (`off-stream'), 158,147 stars were selected within $\pm 0.5$ degrees
of a curve offset by 5 degrees in $\delta$ from the `on-stream' data over
a similar $\alpha$ range.
A Hess diagram in ($(g-r)_0,g_0$) of the
difference between the on and off stream data \citep{nyetal02} was generated and the results
are shown in Figure 1.

Figure 1 shows a clear faint turnoff around $((g-r)_0,g_0) \sim (0.25,19)$ and 
upper main sequence descending to $g_0 \sim 22.5$.  Other features in
Figure 1 include an imperfectly subtracted thick disk residual 
at $(g-r)_0 \sim 0.45$ from $14 < g_0 < 17.5$ and 
a residual from nearby M dwarfs at $(g-r)_0 \sim 1.4$ and $19 < g_0 < 23$.
The stars below the turnoff are concentrated in a relatively narrow band of color and magnitude, 
as expected for a stream localized in distance from
the Sun.  Since the stream varies in distance from 7 to about 10 kpc for the stars in this
figure, the actual width of the main sequence is substantially narrower than demonstrated 
here.  The blue turnoff of $(g-r)_0 \sim 0.25$ suggests a lower metallicity or younger age
than that of the spheroid, which has a turnoff of $(g-r)_0 \sim 0.3$ and $\rm [Fe/H] \sim -1.6$.  
We superimpose with black dots in Figure 1 a fiducial sequence 
from the cluster M92.  The M92 sequence was calculated by starting with the fiducial sequence in
undereddened $r, (g-r)$ from \citet{ann08}, converting it to
$M_{g_0}, (g-r)_0$ using E(B-V) of 0.02 and a distance to M92 of 8.2 kpc, as tabulated by
\citet{h96}, and then shifting it to the approximate distance to the GD-1 stream, which is approximately
9 kpc over this range of RA (distance modulus $m-M = 14.76$).  This low-metallicity cluster, with
$\rm [Fe/H] = -2.3$ \citep{h96}, fits the main sequence and the turnoff reasonably well. 
We do not see the giant branch or the horizontal branch (which should be at $g_0 \sim 15.5$)
in this figure, but given the faintness of the stream that is not unexpected.

We now refine the rough color-magnitude box used to detect the stream in equatorial coordinates.
The refined box is selected to allow for a stream which changes distance
by $\sim 30\%$ over $90^\circ < \alpha < 260^\circ$.  The box is defined as the
union of three selection regions heavily outlined in Figure 1:
A) $0.15 < (g-r)_0 < 0.31, 17.75 < g_0 < 19.7$ (turn off)
B) $0.15 < (g-r)_0 < 0.4, 19.7 <= g_0 < 20.5$ (lower turn off)
and C) $20.5 < g_0 < 22, 17.89 < g_0-6.52(g-r)_0 < 19.52$ (upper main sequence).
All stars with images in the SDSS DR7 Northern Galactic Cap region which
meet these criteria (and have $(u-g)_0 > 0.4$ to exclude quasars), are
selected and plotted in $(\alpha,\delta)$ in the upper panel of Figure 2.
The GD-1 stream is faintly visible, running in an arc from about $(\alpha,
\delta) = (220^\circ,58^\circ) $ through $(164^\circ,48^\circ)$, then
down to $(139^\circ,22^\circ)$ where it
crosses in front of the Sagittarius stream, and then it is not
clearly visible as it is lost in Monoceros and other Galactic halo stars
near $(126^\circ,0^\circ)$.  The lower panel of Figure 2 presents the same
data in a Galactic polar projection.  The GD-1 stream is clearly visible.  There are
several features in the density of stars along the stream of unknown
origin.  They could be the remains of a nearly dissolved progenitor, the result of
interactions between the stream and the potential of the Milky Way's disk and halo,
or unassociated spheroid substructure.

The kinematics of the stream, detailed
below, reveal that it is on a retrograde orbit moving through the
sequence of points in the order just described for the upper 
panel of Figure 2.  Numerous other dwarf galaxies,
streams and halo overdensities are present in Figure 2; these will not
be discussed further here.

\subsection{Spectroscopy}

The SEGUE survey \citep{yetal09}, which is one of three surveys carried out as part of SDSS-II, obtained 
spectra of approximately 240,000 Milky Way stars toward $\sim 200$ sightlines that each
covered seven square degrees of the sky, with an emphasis on obtaining spectra of fainter halo stars. 
While most of SEGUE's 200 observing tiles were randomly distributed across the SDSS imaging 
footprint, a few were placed on streams of known interest, including 
the GD-1 stream.  

All SEGUE spectra were processed through the standard 
SDSS spectroscopic reduction pipelines \citep{stoughton02}, from
which radial velocities accurate to about $10~\rm km~s^{-1}$ for objects
at $g\sim 19.5$ were determined.  In addition, the stellar spectra 
were processed through the SEGUE stellar parameter pipeline (SSPP) \citep{letal08a,letal08b,apetal08} in order to obtain abundance ([Fe/H]), surface gravity (log g),  
and other atmospheric parameter estimates.  

We select from the SDSS-II/SEGUE DR7 database all measured parameters of 
the $12,825$ spectra of stars within $1.3^\circ$ of the GD-1 stream described by Eq. 1.
We further refined the selection
to exclude objects far away from 
the fiducial M92 curve of Figure 1 
by requiring that they meet 
these color magnitude cuts:
$-0.3 < (g-r)_0 < 0.15, 14 < g_0 <17.75$ or
$0.15 < (g-r)_0 < 0.4, 17.75 < g_0 <20.5$ or
$0.4 < (g-r)_0 < 0.5, 15.31 < g_0 <19.14$ or
$0.5 <  (g-r)_0 < 0.59, 14 < g_0 <19.14$ or
$0.59 <  (g-r)_0 < 0.77, 14 < g_0 <17.21$ or
$0.77 <  (g-r)_0 < 1.2, 14 < g_0 <15.31$.
The region bounded by these cuts is outlined with a light line in Figure 1.  
Most of the 4568 remaining spectra are concentrated in $3^\circ$ diameter patches centered on
SEGUE tiles, but some are part of the SDSS-I and SDSS-II Legacy surveys.
These latter surveys targeted nearly the entire SDSS 
footprint spectroscopically, but with few and limited signal-to-noise on stellar targets (since
the SDSS Legacy survey primarily targets galaxy and quasar candidates).

We show in Figure 3 the line-of-sight, Galactic standard of rest velocities, $v_{\rm gsr}$, for each star
in the sample, as a function of Galactic longitude.  We calculate $v_{\rm gsr}$ using:
$v_{\rm gsr} = \rm RV + 10.1~cos\>{\it b}~cos\>{\it l} + 224~cos\>{\it b}~sin\>{\it l} + 6.7~sin\>{\it b}$, where
RV is the heliocentric radial velocity 
in $\rm km~s^{-1}$ and $(l,b)$ are the standard, Sun-centered Galactic 
coordinates of each star.   
A sine curve with amplitude 110 $\rm km~s^{-1}$ traces an approximate
locus of nearby disk stars co-rotating with the Sun.  Spheroid stars occupy a broad range
of $v_{gsr}$ centered at $v_{gsr}=0$.  Seven regions of interest
are marked along the bottom of Figure 3, indicating areas with SEGUE plates, where stars 
identified with the GD-1 stream will be selected.  The positions of these seven regions on
the sky in equatorial coordinates are also indicated with circles 
and numbered in the upper panel of Figure 2.  Regions 5 and 6 were 
specially targeted by SEGUE with a tile directly on locations along the
GD-1 stream.  

From examination of Figure 3, it appears that there is an excess of stars off
the rotating disk locus at $v_{\rm gsr} \sim -90~ \rm km~s^{-1}$ in regions 5 and 6.  To 
confirm that these are in fact GD-1 stream stars, we isolate the stars 
in regions 5 and 6 and plot their velocity histogram in Figure 4.

The distribution in Figure 4 is overlayed with Gaussians for the 
thick disk
(dispersion of $30~\rm km~s^{-1}$ and an offset of $\mu \sim 20 ~\rm km~s^{-1}$),
and inner halo (dispersion of $100~\rm km~s^{-1}$).
A significant peak is detected at $v_{\rm gsr} \sim -82 ~\rm km~s^{-1}$ which we associate with the
GD-1 stream member stars. 

We next examine the metallicity distribution of stars in this velocity peak 
in order to estimate the elemental abundance of the GD-1 stream.  
Later in the paper we will show that the individual $v_{gsr}$ velocities
in regions 5 and 6 are $71 \pm 2$ and $87 \pm 2$ km s$^{-1}$, respectively, so we chose a ``peak"
velocity range of $-97<v_{gsr}<-61$ km s$^{-1}$.  

Figure 5 shows the SSPP
abundance estimates for all stars with good metallicity 
estimates (for a good estimate a turnoff star generally needs to be brighter
than about $g\sim 19$).  Errors on individual stars $\rm [Fe/H]$ are 
approximately 0.3 dex for spectral type F objects with $g < 18.5$.  
The histogram for all abundances of stars in regions 5 and 6 are plotted with 
a light line (1311 stars), those for stars in the velocity peak of Figure 4 are indicated with
a heavy line (115 stars).  
The stars with velocities of the GD-1 stream are heavily biased towards lower metallicity
stars, compared with those of the 
thick disk ($\rm [Fe/H] \sim -0.7$), or inner halo ($\rm [Fe/H] \sim -1.6$). 

We estimate from Figure 4 that about 30 stars in the spectroscopic dataset
are from the GD-1 stream.  To see the metallicitydistribution of the
stars in the GD-1 stream, we subtract a scaled version of the histogram
with the light line from the histogram with the heavy line.  The scaling
factor is (115-30)/(1311-30).  Since the stars in the velocity selected
region contain a smaller fraction of thick disk stars, the
subtracted histogram is oversubtracted at high metallcities, and likely
undersubtracted at spheroid metallcitites.  The mean of the stars in the
shaded region is [Fe/H]=-1.9, but the real metallicity of the stream is
probably somewhat lower than this.
Bins with negative counts do not appear in Figure 5.

We now return to the sample of stars in Figure 3, and select only those of
very low metallicity ($-2.3< \rm [Fe/H] < -1.65$) in order to isolate
stream members from the thick disk and halo field star populations.  
The low metallicity spectra with positions, colors, and magnitudes that make them
candidates for GD-1 stream members are shown in Figure 6.
Several velocity peaks are now clearly separated from the disk and spheroid.
We now examine stars in
each of the seven regions numbered in Figure 6 and determine their
observational properties.

\section{Photometric Distance estimation}

We estimated the distance to the GD-1 stream at the positions of each of the seven
regions that it overlaps using a matched filter algorithm.
We first generated the Hess diagram from SDSS DR7 data from a region about $5^\circ$ wide
in RA and 1$^\circ$ wide in Dec in the vicinity of each plate, centered on the 
polynomial fit to the GD-1 stream. 
Then the Hess diagram of the background was generated from two regions of sky with the same
angular extent on the sky, but offset 1.5$^\circ$ higher and 1.5$^\circ$ lower in declination.
The background Hess diagram (divided by two to correct for the difference in sky area) was 
subtracted from the corresponding Hess diagram on the GD-1 stream.  The subtracted Hess
diagrams are shown in Figure 7.

Since the GD-1 stream is of quite low metallicity, we selected the globular cluster M92
([Fe/H]$\sim -2.3$) to compare with the observations.
We then constructed a M92 filter Hess 
diagram with the similar method to \citet{Grill09}.  We first broaden the M92
fiducial sequence from \citet{ann08} with the SDSS photometric errors.  Because
we do not have a luminosity function for M92 stars, we used the luminosity function of
M13, estimated from SDSS survey counts vs. magnitude for stars away from 
the core of M13, to create the Hess diagram.  
As before, we assume the distance to M92 is 8.2 kpc.

The M92 filter was shifted from -0.5 to 1.5 magnitudes in r in steps of 0.05 mag.
For each shift, we
cross-correlated the M92 filter with the subtracted Hess diagram:
   \begin{equation}
     a({\delta}r)=\sum_{g-r,r}[O(g-r,r)-B(g-r,r)]\cdot F(g-r,r+{\delta}r)
   \end{equation}
and estimated the error of cross-correlation function as:
\begin{equation}
  \sigma(a)=\{\sum_{g-r,r}[O(g-r,r)+B(g-r,r)]\cdot F^2(g-r,r+{\delta}r)\}^{0.5}.
\end{equation}
In the above equations, O and B represent the Hess diagrams of orbit and background segments,
respectively.  F is the value of M92 filter.

Then we define the maximum position, ${\delta}r_m$, of the cross-correlation function to be
 the actual distance modulus to the stream.  Near the maximum, we can estimate the
cross-correlation function by the Taylor expansion:
\begin{equation}
 a({\delta}r)=a({\delta}r_m)+(\frac{da}{d{\delta}r})\mid_{{\delta}r={\delta}r_m}{\delta}r+\frac{1}{2}(\frac{d^2a}{d{\delta}r^2})\mid_{{\delta}r={\delta}r_m} {\delta}r^2+\ldots
\end{equation}
The second term in the right equals zero, since the first derivative is zero at maximum.
We use the above Taylor expansion, using only the lowest order non-zero terms, to estimate
the error in the cross-correlation function:
\begin{equation}
\sigma^2(a)=<a^2({\delta}r)-a^2({\delta}r_m)>=a({\delta}r_m)(\frac{d^2a}{dr^2})\mid_{{\delta}r={\delta}r_m}<{\delta}r>^2.
\end{equation}
Since $<{\delta}r>=\sigma_{{\delta}r}$,
\begin{equation}
\sigma_{{\delta}r}=\sqrt{\frac{\sigma^2(a)}{a({\delta}r_m)\frac{d^2a}{d{\delta}r^2}}}\mid_{{\delta}r={\delta}r_m}
\end{equation}

The distance and corresponding error of each of the seven points along the stream with SEGUE
spectra are listed in Table 1. Figure 8 shows the same data as Figure 7, but overplotted
with the M92 fiducial sequence shifted to the best estimated distance for each sky position.

\section {Notes on Individual Regions}

Figure 9 shows velocity histograms of low metallicity stars from 
Figure 6 within $\pm 1.3^\circ$ of each selected region.  A Gaussian velocity histogram, 
representing a halo distribution with $\sigma = 100 \rm ~km~s^{-1}$ and are normalized
so that the area under the curve equals the number of stars in the histogram, is 
also shown.  The region number is indicated in the upper right corner of each panel,
and the velocity of the peak most likely associated with the GD-1 stream is also
indicated.  
We list below some details
of each region's selection, including the SDSS/SEGUE plates on which 
most of the objects were detected.

In each region with a clear stream detection, the velocity and velocity 
dispersion for the GD-1 stream were computed using an iterative method that
used only stars within one standard deviation of the mean velocity.
We computed the mean and standard deviation of the stars near
the velocity peak.  Then, we selected stars that were within one
standard deviation of the mean and re-computed the mean and standard
deviation.  The standard deviation calculated this way is an underestimate,
since we have removed the tails of the distribution.  We corrected
the standard deviation assuming Gaussian tails.  This process was
repeated until the computed mean and standard deviation matched the
mean and standard deviation used to select the stars in the stream.

Table 2 lists 48 high confidence GD-1 stream members.  The sample
includes the stars in Figure 6 (which are selected 
based on angular distance from the GD-1 stream, photometric color and
magnitude, and metallicity) which have Galactic longitude within
two degrees of the seven plate centers, have velocities near the
measured or expected velocities for the GD-1 stream, and which have
proper motions that are consistent with our GD-1 stream model
(presented in \S 7).  The velocity cuts for each of the seven regions
are: 1) $97<v_{gsr}<118$, 2) $59<v_{gsr}<79$, 3) $32<v_{gsr}<52$,
4) $-17<v_{gsr}<3$, 5) $-78 < v_{gsr} < -58$, 6) $-97 < v_{gsr} < -77$,
and 7) $-166 < v_{gsr} < -146$.  The proper motions expected for
each of the seven regions are listed in Table 1.  The stars in
Table 2 are within two sigma of the expected proper motions, where
one sigma is 3 mas/yr in each of $\mu_l$ and $\mu_b$.
Table 2 includes each objects SDSS-ID number, coordinates, magnitude, colors,
velocity, estimated metallicity, surface gravity and proper motion.
The proper motions ($\mu_l,~\mu_b$) listed here are 
from the USNO-B/SDSS catalog of \citet{munnetal04}, as extracted 
from the DR7 database's `propermotions' table.  Errors on an 
individual $g\sim 18$ star's measurement are about $3 \rm ~mas~yr^{-1}$ on
each coordinate.

\subsection {Primary Kinematic Regions}


Regions 4, 5, and 6 were specifically targeted by SEGUE with plates
along the GD-1 stream.   These regions, along with region 1, targeted
by the SDSS Legacy survey, constitute the 4 regions along the
stream with spectroscopy used to fit a model orbit for GD-1.

{\it Region 4, Plates 2889 and 2914:}
The GD-1 stars are well sampled in this plate pair; there is a
clearly detected peak of (primarily) F turnoff stars 
at $v_{gsr}= -7 \pm 1$ km s$^{-1}$ with $\sigma = 3.9$ km s$^{-1}$, in the
magnitude range $17.5 < g_0 < 19.5$. The 
metallicity of stars in the velocity peak is about 
$\rm [Fe/H] \sim -2.2$.
There are two BHBs in this sample near $g\sim 14.9$, corresponding
to a distance of 7.2 kpc \citep{sirko04}.  The distance to this
region derived in Section 3 from the turnoff photometry is $7.5\rm \pm 0.33 $kpc,
in good agreement with the BHB magnitudes.  
The velocities
of the two BHBs are -6 and -3 $\rm km~s^{-1}$, in excellent
agreement with the average of the turnoff star velocities, as are the
metallicities ([Fe/H] = -2.0 and -2.1, respectively). 

{\it Region 5, Plates 2557 and 2567:}
The plate spans the full width of the GD-1 stream.  There is a
strong peak in the velocity distribution. 
This peak has $v_{gsr} = -71 \pm 2 ~\rm km~s^{-1}$ with $\sigma=5.3$ 
km s$^{-1}$.  The 
average $g$ magnitude in this range is $g\sim 19$, which is 
consistent with the distance estimation from color magnitude turnoff 
fitting.  The metallicity distribution of F turn-off 
stars in the peak shows that [Fe/H] of this stream is about -2.05.
The distance to this piece of the stream is $8.0\pm 0.53~\rm kpc$.

{\it Region 6, Plates 2390 and 2410:}
A strong and narrow peak is detected in Figure 9, at
$v_{gsr}=-87 \pm 2\rm ~km~s^{-1}$, $\sigma=9$ km s$^{-1}$ and $g_0=18.82$. 
[Fe/H] distributions show the metallicity peaks at about
-2.05, which is consistent with the [Fe/H] peak find in plate 2567 and
slightly higher than plate 2914, this indicates that they are from the
same stream.    A distance estimate puts stars in this region of the stream 
at $8.8\pm 0.75~\rm kpc$ from
the sun.


{\it Region 1, Plate 1154:}
This is a special SDSS legacy plate, rather than a SEGUE plate, but since it
is at low Galactic latitude it
does have a large number of stellar candidates.
Its important GD-1 stream candidate stars cluster at 
velocity $v_{gsr} = 108 \pm 5 ~\rm km~s^{-1}$, with $\sigma_v = 11$ km s$^{-1}$,
and (lower S/N) metallicity [Fe/H] $= -2\pm 0.3$, which anchors the 
stream orbit away from regions 4,5 and 6.
With larger errors, the distance to this stream 
here is $10.4\pm 1.2~\rm kpc$.
There is a second peak (3 stars) at $v_{\rm gsr} = -40 ~\rm km~s^{-1}$.
We discount this second peak as being associated with GD-1, since two of 
its three members have [Fe/H] = -1.8, significantly higher than 
the average for other stream peak members. This
secondary peak could be related to the Anti-Center stream noted by \citet{gcm08}, as the radial velocity of these stars is $\rm RV\sim +112~\rm km~s^{-1}$, is close
to their value for ACS-C (see lower panel of Figure 1 and Figure 2 of that work).  

\subsection {Other Regions}

The following 3 regions were not involved in the model fit (see below),
but as there are SEGUE spectra available here along the orbit defined
by the imaging arc defined in Equation 1, we examine these
SEGUE plates for possible stream members.

{\it Region 2, Plates 1760, 2433, 2667 and 2671:}
This plate pair was targeted by SEGUE as it contains the open solar-metallicity
cluster M67.  The Sagittarius stream also passes near by, along with the
Anti-center stream. By chance, the GD-1 stream arc appears to pass within 
1.3$^\circ$ of the plate center, and several very low metallicity turnoff stars
are detected with average $v_{gsr}=69~\rm km~s^{-1}$.  The distance to the stream
is $6.5\pm 0.64~\rm kpc$.

{\it Region 3, Plates 2304 and 2319:}
Several stars with metallicity have velocities in a broad excess near
$v_{gsr} = 42 ~\rm km~s^{-1}$, however there is not a significant candidate 
peak here.  
A peak at $v_{\rm gsr} \sim -105 \rm ~km~s^{-1}$ has stars with $<[\rm Fe/H]> \sim -1.8$, and is not a viable stream maximum.  The distance from section 3 is estimated at
$7.0\pm 0.36~\rm kpc$.
 
{\it Region 7, Plate 2539 and 2547:}
This plate pair is on the fitted arc of Equation 1, but the stream 
becomes too tenuous to be defined.  We do not see 
a velocity peak here, but following the trend in the peaks of points 1-6 
there may be two or three stars (above background)  at
$v_{gsr} = -156\pm 10~\rm km~s^{-1}$, with the correct metallicity 
and proper motion to be associated with
GD-1.  We do not place as strong a confidence in the GD-1 
membership of stars in this peak as the other 6 regions.  A very 
uncertain distance estimate to the stream here, is 
about $d=9.9\pm 1.2~\rm kpc$ from the Sun.  Orbit fits, which were not fit
to data in region 7, confirm that this is the correct velocity to be
looking for stream stars, but that the distance is somewhat underestimated.

We note that the estimated distances to individual regions are in very good
agreement with those quoted by \citet{gd06}, indicating that the results
are somewhat robust to details of the method and potential parameters.

\section{The Observed Stream Properties}

The observed properties of the stars in the seven GD-1 stream candidate
regions are summarized in Table 1, where we list region number ($N$);
Equatorial coordinates, $(\alpha,\delta)$; the corresponding Galactic
coordinates ($l,b$) with errors (we use $\delta$ to denote a measured
error, to distinguish it from the intrinsic dispersion, which
we denote with the symbol $\sigma$); the 
average Galactocentric standard
of rest velocity and heliocentric radial velocity with an error, and the
velocity dispersion.  The velocity mean and dispersion were calculated as 
described in \S 3.  The tabulated intrinsic dispersions are upper limits to the
actual velocity dispersion of the stream; since they are similar in size
to the velocity errors for each individual spectrum, the measurement is
consistent with an intrinsic velocity distribution of zero.  The 
error in the mean is the velocity dispersion
divided by the square root of the number of stars used to compute it.
Regions 2, 3, and 7 
do not have clear, narrow peaks in the velocity distributions and therefore 
were not used to fit the orbit, though Figure 6 shows there are excess stars at
about the right velocities.

Table 1 lists the computed Galactic $X, Y,
Z$ positions (with respect to a right handed coordinate system with the Sun at
(-8,0,0) kpc) for each stream region and a distance to the stream at
that region, again with errors.  For the four regions numbered 1,4,5
and 6, the velocity and position accuracies are highly significant,
and there is high confidence of the GD-1 stream membership of stars 
highlighted in the corresponding boxes of Figure 6.  

Figure 10 highlights this confidence by showing a color-magnitude
diagram of all high-confidence spectroscopic GD-1 stream candidates 
in regions 1-7, shifted to a standard distance of 9 kpc based on
the photometric Hess diagram fitting.
These objects are also proper motion selected, in that only objects
within $\pm 2 \sigma$ of the proper motion of the best fit model are
kept.
Superimposed over the spectral candidates is the
M92 fiducial sequence of \citet{ann08} shifted to a distance
modulus of $m-M = 14.76$, identical to that used in Figure 1.
Distances to the other regions
were estimated as described in section 3.

We now calculate our best metallicity estimate for GD-1 by selecting only the
48 stars with spectra in Figure 10.  We note that these stars 
were pre-cut on metallicity at an earlier stage (Figure 6)
to have $-2.3 < \rm [Fe/H] < -1.65$.
A histogram with bins similar to the measurement error
yields a GD-1 stream metallicity of [Fe/H]$=-2.1\pm0.1$ dex 
with a dispersion of $\sigma = 0.3$ dex (essentially the measurement error).  
In addition to these statistical errors, there may be systematic errors in 
the metallicity determinations from SDSS DR7 of $\sim 0.2$ dex \citep{apetal08}.



\section{Orbit Fitting}

We now use the data listed in Table 1 to fit an orbit for the stream, assuming a fixed
Galactic potential.  \citet{gd06} postulated the progenitor of this stream is a
globular cluster because it has a narrow width in the sky. As a cluster orbits the Galaxy, stars farther from the progenitor will depart from the orbit due to dynamical friction and scattering of the stream stars. 
Because the progenitor
is presumed to be a compact object with a few km s$^{-1}$ velocity dispersion, it is reasonable
to assume that the stars in the tidal stream lie approximately on the orbit of the
globular cluster \citep{odenkirchen03}. Dwarf galaxies, on the other hand, will experience larger spatial dispersions because they have larger dispersions in their energies. 
Therefore, in this paper, we fit the orbit to the positions and velocities
of the stars in the tidal stream. 

\subsection{Galactic Model}

The Galactic potential model used in this work follows directly from \citet{law05}. 
We use a 3 component potential - disk, bulge, and halo, of the form given in 
Equations (\ref{disk}), (\ref{bulge}), and (\ref{halo}).

\begin{equation}
\Phi_{disk} = -\alpha \frac{G M_{disk}}{\sqrt{R_{c}^2 + (a + \sqrt{Z^2 + b^2})^2}}
\label{disk}
\end{equation}

\begin{equation}
\Phi_{bulge} = - \frac{G M_{bulge}}{r+c}
\label{bulge}
\end{equation}

\begin{equation}
\Phi_{halo} = v^2_{halo} \ln \left({R_{c}^2 + \frac{Z^2}{q^2} + d^2}\right) 
\label{halo}
\end{equation}
In these potentials, $R_{c}^2 = X^2 + Y^2$ and $r^2 = X^2 + Y^2 + Z^2$, where $(X,Y,Z)$ are 
Galactocentric Cartesian coordinates. We adopt the Sun - Galactic center 
distance so that $X_{Sun} = -8.0 \rm ~kpc$. 
The symbol $R_c$ is the cylindrical radius
from the center of the Galaxy, whereas $R$ refers to the 
Sun-centered distance to an arbitrary point along the stream. In these potentials, $M_{disk}$ and $M_{bulge}$ are the masses of the disk and bulge, respectively. The spatial extent of the potentials are determined by $a$, the disk scale length, $b$ the disk scale height, $c$ the bulge scale radius, and $d$ the dark matter halo scale length. Additionally, $v_{halo}$ describes the dark matter halo dispersion speed (which is related to the total dark matter halo mass), and $q$ represents the dark matter halo flattening in the $Z$ direction.  We found that the 
parameters of the Galactic potential were not well constrained by the path of the 
tidal stream, so these
parameters (Table \ref{parameters}) were held constant, with the same values used
by \citet{law05}.

In this work, we will fit four orbital parameters, given in 
Table \ref{fitpars}. $(R_5,v_{x,5},v_{y,5},v_{z,5})$ are
the distance (from the Sun) to, and velocities of, region 5.  Given a
Galactic potential, these parameters fully specify the GC orbit.
We then evolve the test particle forward and backward from the starting location, 
using the $mkorbit$ and $orbint$ tools in the NEMO Stellar Dynamics Toolbox \citep{teuben}. We  convert the resulting orbit into $(l,b)$ and perform the 
goodness of fit calculation.
%

To find reasonable initial values for these parameters, we imagine placing a 
test particle in region 5 at  $(l,b,R_5) = (172.3^\circ, 57.43^\circ,8.0$ kpc). 
We then construct a vector between the $(l,b,R_5) = (172.3^\circ, 57.43^\circ,8.0\rm ~kpc)$ and 
$(l,b,R_6) = (161.95^\circ,59.12^\circ, 8.8~\rm kpc)$ points. This gives us the direction of the 
total velocity, because we are assuming the orbit passes through both of these points. The 
principle initial values for the parameters in region 5 are 
$(R_5,v_{x,5},v_{y,5},v_{z,5})$ = (8.0 kpc, -94 km/s, -285 km/s, -104 km/s).  In practice
we start searching for the best parameters in a range of values near these approximate
values for the orbital parameters. The parameter selection ranges are given in Table \ref{fitpars}.


\subsection{Goodness-of-Fit}
In order to find the best-fit orbit to the data given in Table 1, it is necessary to define 
a goodness-of-fit metric, and search the relevant parameters for the minimum
 value of this metric. The metric for an orbit of this type involves three 
important factors: the orbit passing through the plate locations in the sky, 
having the proper velocities at these locations, and having the correct distances. 
In order to consider all of these factors, we define three chi-squared values, 
one for each of the relevant variables, and simply sum them to create 
the total goodness of fit metric. Specifically,

\begin{equation}
\chi^2_{b} = \sum_{i} \left(\frac{b_{model,i} - b_{data,i}}{\sigma_b}\right)^2,
\label{bchisq}
\end{equation}

\begin{equation}
\chi^2_{rv} = \sum_{i} \left(\frac{rv_{model,i} - rv_{data,i}}{\sigma_{rv,i}}\right)^2,
\label{vchisq}
\end{equation}

\begin{equation}
\chi^2_{R} = \sum_{i} \left(\frac{R_{model,i} - R_{data,i}}{\sigma_{R,i}}\right)^2, {\rm ~and}
\label{Rchisq}
\end{equation}

\begin{equation}
\chi^2 = \frac{1}{\eta} \left(\chi^2_{b} + \chi^2_{v} + \chi^2_{R}\right),
\label{chisq}
\end{equation}
where $\eta = N - n -1$, $N$ being the number of data points, 
and $n$ being the number of parameters.

To calculate these $\chi^2$ values, we calculate a model orbit using the 
selected parameters. We search the orbit for the $l$ values from the plates, 
and use the associated values of $b$, $v_{gsr}$, and $R$ to compute $\chi^2$.  

\subsection{Gradient Descent}

We now optimize the orbital parameters so that $\chi^2$ is minimized.
To do this, we choose an initial 
set of parameters, calculate an orbit using the NEMO Stellar Dynamics Toolbox \citep{teuben}, 
and calculate $\chi^2$. We then use a gradient descent method to adjust the 
parameters to new values, generate a new orbit, and recalculate $\chi^2$. This process 
is continued until the true minimum value of $\chi^2$ is found, and the associated parameters 
describe the best fit orbit. 

Let the vector $\vec{Q} = (R_5,v_{x,5},v_{y,5},v_{z,5})$ describe the parameters.
For each $\vec{Q}$ there is an associated $\chi^2 (\vec{Q})$. We choose an initial 
set of parameters $\vec{Q_0}$, and find $\chi^2 (\vec{Q_0})$. We then iterate the 
parameters by Equation (\ref{gradientdescent}).

\begin{equation}
Q_{i,new} = Q_{i,old} - h_i \Lambda \vec{\nabla}_i \chi^2(\vec{Q}_{old})
\label{gradientdescent}
\end{equation}

We calculate the gradient using a finite difference method, 
shown in Equation (\ref{finitedifference}).

\begin{equation}
\nabla_i \left.\chi^2(\vec{Q}) \approx \frac{\chi^2(Q_i + h_i) - 
\chi^2(Q_i - h_i)}{2h_i} \right|_{\hbox{all other $Q_k$ fixed}}
\label{finitedifference}
\end{equation}

Different values of $h_i$ are used because the parameters are on 
different scales, it would not be appropriate to use the same step size for 
them all. The step sizes for the parameters are given in Table \ref{fitpars}.

$\Lambda$ is a variable-learning parameter. It initially begins at $\Lambda = 1$, and 
if the new value of $\chi^2(\vec{Q})$ is smaller than the old, then $\Lambda$ is 
multiplied by $1.03$, if not, it is multiplied by $0.80$. The purpose of this is 
to ensure if a minimum is being found, then it is found faster than with a 
constant-learning parameter. We also multiply it by the associated $h_i$ value to make the 
step size appropriate for the parameter being considered.

\subsection{Error Estimation}

The uncertainties of the best-fit parameters can be estimated from 
the shape of the $\chi^2$ surface at its minimum. To do this, we follow 
the method outlined by \citet{cole08}. We construct a matrix $\mathbf{V}$ of 
second partial derivatives of the $\chi^2$ surface, evaluated at the minimum 
found by the gradient descent. The error estimate for the $i^{th}$ parameter 
is $\sigma_i = \sqrt{2V_{ii}}$. The $\mathbf{V}$ matrix is defined as

\begin{equation}
\mathbf{V} \equiv \mathbf{H}^{-1}.
\label {Vdef}
\end{equation}

The matrix $\mathbf{H}$ is the Hessian Matrix, whose elements are given by

\begin{eqnarray}
H_{ij} & = & \frac{H^1_{ij} - H^2_{ij} - H^3_{ij} + H^4_{ij}}{4 h_i h_j} \hbox{, where,} \\
H^1_{ij} & = & \left.\chi^2(Q_j + h_j, Q_i + h_i)\right|_{\hbox{all other $Q_k$ fixed}} \nonumber \\
H^2_{ij} & = & \left.\chi^2(Q_j - h_j, Q_i + h_i)\right|_{\hbox{all other $Q_k$ fixed}} \nonumber \\
H^3_{ij} & = & \left.\chi^2(Q_j + h_j, Q_i - h_i)\right|_{\hbox{all other $Q_k$ fixed}} \nonumber \\
H^4_{ij} & = & \left.\chi^2(Q_j - h_j, Q_i - h_i)\right|_{\hbox{all other $Q_k$ fixed}} \nonumber \\
\label {Hdef}
\end{eqnarray}

\section{Results and Discussion}
We select five initial sets of parameters by randomly choosing values within the ranges 
given in Table \ref{fitpars}. We perform the gradient descent to reach the best fit 
parameters.  We then estimate the parameter errors using the 
Hessian method outlined above. The best-fit parameters and their errors are 
$(R_5,v_{x,5},v_{y,5},v_{z,5}) = (8.4 \pm 0.8 \rm ~kpc, -89 \pm 2 \rm ~km/s, -236 \pm 6 \rm ~km/s, 
-115 \pm 3 \rm ~km/s)$. The chi-squared of this fit is 2.07.
The negative velocities indicate a retrograde orbit. The perigalacticon for
this orbit is located $r = 14.43\pm 0.5$ kpc from the Galactic
center at $(l,b) = (158^\circ,60^\circ)$, near region 6.  The space velocity
of stars in the model at this position is 276 $\rm km~s^{-1}$.   The apogalacticon is at $r = 28.7\pm 2$ kpc from the Galactic center, toward $(l,b) = (306^\circ,-35^\circ)$, though we do not observe this direction on the sky.  All errors
are $1\sigma$.

These orbital parameters are fairly insensitive to the choices of parameters in the Galactic
potential.  In particular, they are good estimates of the orbital parameters for a wide range
of $q$ and $d$.  To investigate whether we could determine anything about the shape of the
dark matter halo from this tidal tail, we performed parameter sweeps on $q$ and $d$ while keeping 
the kinematic parameters constant (Figure 11).  We see that very low flattenings are 
inconsistent with the data, but a wide range of $q$ and $d$ are allowed.

The potential assumed in fitting the orbit is a standard logarithmic flat-rotation curve dark matter
halo plus a stellar disk.  Since the GD-1 stream approaches within 15 kpc of the Galactic 
center, the effects of the massive disk are felt by the orbit, and increasing the 
relative mass of the disk vs. the halo can mimic the effects of a flattened halo. At perigalacticon, the disk exerts twice as much gravitational force as the halo. 
Given this, it is not surprising that the stream orbit depends minimally on the halo parameters. 
More models and constraints, from this and other streams, are clearly needed to constrain the shape of the dark matter halo.

The upper panel of Figure 12 shows the orbit in $(l,b)$ with the stream locations shown. The model 
prediction is in very good agreement with the experimental observations. 
The middle and lower panels of Figure 11 show the orbit in $l$ versus $v_{\rm gsr}$ 
and $l$ versus distance from the Sun. We also see good agreement 
with the experimental observation.  Figure 13 shows the orbit projected into
the three planes of Galactic coordinates $(X,Y,Z)$.  We deduce an orbital
eccentricity $e = 0.33\pm0.02$ (one sigma error) and an inclination to the Galactic plane of $i\sim 35\pm 5^\circ$.
Arrows show the relative direction of the stream's retrograde motion compared to the Milky Way.

The final columns of Table 1 show the predicted proper motions ($\mu_l,~\mu_b$) for stars
in the stream at each region 1-7 based on the distances in Table 1.  These predictions may be compared with actual observed proper motions for stream star candidates in Table 2 at
each region.  In general the agreement is quite good for regions 2-6, given proper motion
errors of $1\sigma = 3 \rm ~mas~yr^{-1}$ in each coordinate.  Spectral candidates more than
$2\sigma$ away were excluded from Figure 10, dropping about 20\% of the candidates,
leaving a generally good fit to a shifted M92 fiducial sequence for these regions.
Region 7 had fewer good proper motion matches, and it is possible that we are not seeing
GD-1 stream candidates here. 

Figure 14 shows the photometrically chosen stars with proper motions available near regions 1,4,5 and 6,
along with an equivalent set of field stars (chosen 5 degrees away) for 
comparison.  There's a clear excess of `on-stream' stars in the lower right 
quadrant of each region on-stream -- these are likely stream members.  
The estimated tangential velocities (relative to the Sun) are given.

To search for a possible progenitor, we selected all Milky Way globular clusters 
from \citet{h96} that had metallicities in the range $-2.5 <$[Fe/H]$<-1.5$.  Only seven of 
these globular clusters (Terzan 8, Arp2, NGC 6809, NGC 6749, NGC 6341, NGC 6681, and NGC 6752) 
are within $5^\circ$ of the GD-1 orbit.  Additionally, we considered NGC 2298, which is $5.5^\circ$ from the orbit, and has a metallicity of $\rm [Fe/H] = -1.85$. We compared the positions and velocities of these
globular clusters with an orbital path that extends all the way around the Milky Way.  To
create a stream of this length would require a globular cluster to orbit the Milky Way for
on the order of gigayears, with the length depending on the concentration of the
progenitor, as well as the shape and location of the progenitor's orbit.  Of the eight globular clusters, NGC 6809, NGC 6749, and NGC 6752 are
ruled out because their distances are more than a factor of two different from the distance
to the orbit.  The remaining five clusters had radial velocities that are inconsistent with
the predicted orbit by more than 50 km/s. 
We therefore conclude that the Milky Way globular 
cluster catalog published by \citet{h96} does not contain the progenitor of this stream.
 
\section{Conclusions}


We use spectroscopic kinematic and abundance information to isolate stars in the
GD-1 stream, and use the positions and velocities of those stars to derive orbital parameters for
its orbit.  The GD-1 stream is moving very rapidly on a retrograde orbit around the
Milky Way.  In the region of the orbit which is detected, it has a distance of about 7-11 kpc 
from the  Sun. The stream's orbit takes it to apogalactic distances of $28.75\pm 2$ kpc, and it has a 
perigalacticon of $14.43\pm 0.5$ kpc, implying an eccentricity of $0.33\pm 0.02$.  The inclination to the Galactic 
plane is about $i\sim 35^\circ\pm 5$.  The metallicity of the stream 
is [Fe/H]$ \sim -2.1\pm 0.1$ plus systematic errors of a few tenths dex. 
None of the known globular clusters in the Milky Way have positions, radial velocities, and
metallicities that are consistent with being the progenitor of the GD-1 stream.

The consistency between the proper motions of these stream candidates and our best fit model
gives us further confidence that we have identified stream members and that our
model accurately represents the path on the sky of the stream stars.  While we claim
only consistency here between the proper motion data and our model, we note that more 
detailed fits to the proper motion (in addition to the radial velocities) for such nearby 
streams can be a crucial tool in constraining the halo potential shape and other parameters.  

We acknowledge a careful reading and several important suggestions from the anonymous referee 
which significantly improved the observational analysis section of this paper.
B.A.W and H.J.N. acknowledge support from the National Science Foundation, grant AST 06-06618. We 
gratefully acknowledge Peter Teuben for many useful NEMO discussions and Lee Newberg for assisting
us with calculating the measurement errors.  We also acknowledge useful discussion with Linda Sparke 
regarding streams in potentials.  Z.H.T. acknowledges support from the National Natural Science 
Foundation of China under Grant No. 10778711.  T.C.B. has received partial support for this work 
from grants PHY 02-16783 and PHY 08-22648: Physics Frontiers Center / Joint Institute for Nuclear 
Astrophysics (JINA), awarded by the U.S. National Science Foundation.

Funding for the SDSS and SDSS-II has been provided by the Alfred
P. Sloan Foundation, the Participating Institutions, the National
Science Foundation, the U.S. Department of Energy, the National
Aeronautics and Space Administration, the Japanese Monbukagakusho, the
Max Planck Society, and the Higher Education Funding Council for
England. The SDSS Web Site is http://www.sdss.org/.

The SDSS is managed by the Astrophysical Research Consortium for the
Participating Institutions. The Participating Institutions are the
American Museum of Natural History, Astrophysical Institute Potsdam,
University of Basel, Cambridge University, Case Western Reserve University, 
University of Chicago, Drexel University, Fermilab, the
Institute for Advanced Study, the Japan Participation Group, Johns
Hopkins University, the Joint Institute for Nuclear Astrophysics, the
Kavli Institute for Particle Astrophysics and Cosmology, the Korean
Scientist Group, the Chinese Academy of Sciences (LAMOST), Los Alamos
National Laboratory, the Max-Planck-Institute for Astronomy (MPIA),
the Max-Planck-Institute for Astrophysics (MPA), New Mexico State
University, Ohio State University, University of Pittsburgh,
University of Portsmouth, Princeton University, the United States
Naval Observatory, and the University of Washington.

\clearpage

\figcaption{The GD-1 stream Hess diagram} {
Hess diagram of on-stream minus off-stream (taken from 5 degrees lower 
$\delta$ than the on-stream stars) star counts.  Darker areas of the
figure have higher stellar density.  Thick disk stars and M dwarfs
did not subtract perfectly between on and off regions.  The 
blue turnoff at $((g-r)_0,g_0) = (0.25,19)$ indicates that the GD-1 stream 
is relatively metal poor.  A fiducial sequence for globular cluster M92
from \citet{clemetal08}, shifted to $m-M = 14.76$, 
is overlaid with black dots.  The region of color-magnitude space used 
to select the photometry of stars in Figure 2 is outlined in heavy
black lines.  The region of color-magnitude space that is used to select
the stars plotted in Figure 3 is outlined in light black lines.  
\label{figure1}
}

\figcaption{The GD-1 stream photometry} {
The upper panel shows a Hess diagram of stars in the SDSS footprint which
are within the color and magnitude boxes thickly outlined in Figure 1.
The GD-1 stream arcs faintly from 
$(\alpha,\delta) = (220^\circ,58^\circ) \rm ~to~ (126^\circ,0^\circ)$.
Seven regions where spectroscopy of GD-1 stream star candidates have been obtained are numbered.
Squares indicate regions where the stream was clearly found in velocity, and circles indicate
additional plates that may contain stream stars.
The lower panel shows the GD-1 stream in Galactic polar coordinates.
The color-magnitude cuts are similar to the above, but plotted in
Galactic polar coordinates as labeled. Note the stream's path on the sky.
\label{figure2}
}


\figcaption{Velocities of stars near the stream} {
We plot the line-of-sight, Galactocentric standard of rest velocity versus Galactic
longitude for SEGUE and SDSS stars for which we have spectra, and which are close in 
color-magnitude space to the 
fiducial sequence of M92 (as indicated by the narrow dark outlines in Figure 1), which are 
close (within $1.3^\circ$) in projected distance to the GD-1 stream shown in Figure 2.
A sine curve with amplitude $110 \rm ~km~s^{-1}$ is shown to indicate the locus of
stars rotating with the Sun about the Galactic center.  Stars in the halo will have velocities
centered on $v_{gsr} = 0$ and a large $\sigma \rm \sim 100~km~s^{-1}$ dispersion.
Regions where GD-1 stream candidates are followed up on are numbered 1-7.
Note in particular the groups of stars at $v_{gsr} \sim \rm -90 ~km~s^{-1}$ in regions 5 and 6.
\label{figure3}
}

\figcaption{Histogram of $v_{\rm gsr}$ towards regions 5, 6} {
	All stars with spectra from Figure 3
towards regions 5, 6 of Figure 2 are
histogrammed in Galactocentric velocity.
Gaussians are overlaid representing a
thick disk and halo distribution toward this direction.  
The candidate GD-1
stream stars at about $v_{\rm gsr} \sim -82 ~\rm km~s^{-1}$
cannot be explained by either a halo or thick disk distribution.
\label{figure4}	
}

\figcaption{Metallicity distribution of region 5,6 stars} {
Histogram of SSPP metallicities for all stars (light line) and for stars with
$-97 < v_{gsr} < -61$ km s$^{-1}$ (heavy line).
The velocity-selected peak has an excess of
stars with metallicity lower than that of the halo.  The hashed line indicates a correction
to the heavy line for interlopers at other velocities which have the same
metallicities as candidate stream stars.  This figure suggests
that the peak metallicity of the GD-1 stars is lower than [Fe/H]=-1.9.
\label{figure5} 
}

\figcaption{Velocity distribution of low metallicity stars near the stream} {
The subset of stars from Figure 3 with SSPP
metallicities $-2.3 < \rm [Fe/H] < -1.65$ are presented.  Now the counter-rotating 
GD-1 stream stars stand out more clearly
against their field contaminants compared with Figure 3. Regions 1,4,5 and 6 
have clear peaks; the mean and the error on the mean for these
peaks are indicated by the position and height of the rectangles at these
four longitude locations.  The circles in regions 2, 3, and 7 indicate the area
through which the stream should pass if our model is correct. All areas except region 7 
seem to have an excess of stars with the expected stream velocities.
\label{figure6}
}

\figcaption{Background Subtracted Hess diagrams along the stream} {
At each region 1-7, stars with $(u-g)_0 > 0.4$ within $0.5^\circ$ of the cubic described
by equation 1 are selected.  A similar background set of stars, obtained
from a region $1.5^\circ$ away from the cubic, was selected for each of the
seven regions.  The Hess diagrams show the stream star counts minus the
background star counts, as a function of $g_0$ and $(g-r)_0$.
Light (white) regions indicate an excess of stars along the stream.
\label{figurehessnolabels}
}

\figcaption{Hess diagrams with fits} {
The diagrams from the previous figure are fitted with M92 fiducial loci, shown as (green) crosses, shifted up
and down in magnitude until a maximum correlation is obtained (see text).
Candidate stream stars with spectra are indicated as (red) diamonds.
The distance to each region along the stream is then derived from the best fitting shifted M92 fiducial.
\label{figurehesslabels}
}

\figcaption{Velocity selection of stream candidates} {
Velocity histograms in each GD-1 stream
candidate regions 1-7, with SSPP metallicities $-2.3 < \rm [Fe/H] < -1.65$.  A Gaussian with $\sigma = ~100 \rm ~km~s^{-1}$
is overlaid.  The $v_{\rm gsr}$ of the expected GD-1 peak is noted in each panel.
}

\figcaption{Stars with spectra in the stream} {
All stars in regions 1-7 with SEGUE spectra which meet the
metallicity, color, magnitude, velocity and proper motion cuts as described in the text are
plotted with colored points as indicated in the legend, along with estimated distance
to each set of points.  Each set of points was shifted 
to the reference distance of 9 kpc (of Figure 1), and overlaid with
a M92 fiducial locus, shifted to the same distance moduli.
Note the two (blue) points at $g_0 \sim 15.3$, which
are actually at $g_0 \sim 14.9$ before shifting. These are candidate BHB 
stars in the GD-1 stream, at a distance of 7.5 kpc from the Sun in region 4.  
}

\figcaption{Parameter Sweep of $q$ and $d$} {
Fixing the orbital parameters at their best fit values, derived assuming that 
$q=1$ and $d=12$, we varied $d$ and $q$ and measured $\chi^2$.  
This is justified by our observation that the best fit GD-1 orbital
parameters are fairly similar for a large range of assumed $q$ values.  
The normalization of this
$\chi^2$ surface in $d,q$ was chosen so that the $\chi^2$ at $(d,q)=(12,1)$ is 2, to
match the optimization presented in this paper.  The dark squares in the
figure show the locus of lowest $\chi^2$ in this two dimensional surface.  We see that
very low values of $q$ are not favored, while there is almost no constraint on how prolate
the dark matter distribution can be.  The values of $q$ and $d$ are coupled.  The best fit
$d$ for a given $q$ shifts somewhat with small changes to the orbital parameters, so this
plot cannot be used to definitively pin down the $d$ value, even though the $\chi^2$
surface seems narrow in that dimension.
\label{sweep}
}

\figcaption{Best Fit Orbit} {
Upper panel: Galactic coordinates.  The best fit orbit (with fixed $q,d$) to the data in 
the four regions 1,4,5 and 6.  
Region 1 is leftmost, with $l=224.47^\circ$.  Galactic ($l,b$) 
for all seven regions described in the text are plotted as crosses.
Middle panel, plotting Galactic longitude
$l$ vs. Galactocentric radial velocity $v_{gsr}$ for the best fit model
and data.   Regions 1,4,5 and 6 (left-to-right) have the smallest error bars.
Lower panel, plotting Galactic longitude $l$ vs. Sun-centered
distance to the stream. The errors on regions 4, 5 and 6 are small; the
error bars on the other region data points are limited by how well
the position of the stream turnoff (minus a background field) can be 
identified in a color-magnitude Hess density diagram of stream stars.
The distance error bars are largest for region 7 (rightmost), 
where the stream is confusion limited.
\label{figure12}
}

\figcaption{Best Fit Orbit: $X,Y,Z$} {
Shown in heavy black line is the best fit orbit to 
the four regions 1,4,5 and 6 in Galactic rectangular 
coordinates $(X,Y,Z)$ in each of the cardinal projections.
The coordinate system is a right-handed system with the Sun at
at (-8,0,0) kpc and the Galactic center at the origin.
The coordinate and turnoff magnitude data from the seven regions is
converted to $(X,Y,Z)$ assuming an absolute turnoff F star magnitude of
$M_g = +4.2$.  The seven points are indicated with error bars from
distance error estimates.  Regions 1 and 7 are indicated in each panel,
with the other regions falling in order at intermediate positions.
The units of each axis is kpc.   An arrow originating at the Sun 
indicates the direction of Galactic rotation.
The arrows associated with the stream indicate the retrograde 
direction of motion of GD-1 stream stars. The space velocity of the stream
at perigalacticon is approximately 276 $\rm km~s^{-1}$.
\label{figure13}
}

\figcaption{Proper Motion consistency} {
For the four regions N=1,4,5,6 where
we have fitted the orbit, we select stars within $0.3^\circ$ of
a cubic similar to equation 1 which have measured USNO-B proper motions
from the DR7 database.  For reference, we select similar sets of field objects
offset by 5 degrees in declination from the on-stream objects.
We then sub-select stars with colors and magnitudes of stream turnoff candidates 
using the the heavily outlined box of Figure 1, and plot the $\mu_{l}$ vs $\mu_{b}$ of the
on-stream (black dots) and off-stream (open circles) for each selected region.
There is a clear excess of on-stream points extending to the lower right in each region.
We superimpose crosses representing the point where the best fit model (Table 1) crosses 
the stream, and show the derived tangential velocity in the Figure.  The cross always falls in
the same quadrant with the excess of proper motion points.  Typical errors
on each point are $3~ \rm mas~yr^{-1}$ in each direction.  The upper and right
axes in each figure convert the observed proper motions to 
tangential velocities,
assuming the distance to the stars is the distance to the fitted orbit
for that particular GD-1 stream region.
\label{figurepropermotion}
}

\clearpage

{
\begin{deluxetable}{rrrrrrrrrrrrrrrrrr}
\rotate
\tabletypesize{\scriptsize} \tablecolumns{18} \footnotesize
\tablecaption{GD-1 Stream Detections} \tablewidth{0pt}
\tablehead{
\colhead{N} & \colhead{$\alpha$} & \colhead{$\delta$}  & \colhead{$l$} & \colhead{$\delta l$} & \colhead{$b$} & \colhead{$\delta b$} & \colhead{$\rm v_{gsr}$} & \colhead{RV} & \colhead{$\delta v$} & \colhead{$\sigma_v$} & \colhead{X}& \colhead{Y} & \colhead{Z} & \colhead{d}& \colhead{$\delta d$} & \colhead{$\mu_l$} &\colhead{$\mu_b$} \\
\colhead{}& \colhead{$^\circ$} & \colhead{$^\circ$}  & \colhead{$^\circ$} & \colhead{$^\circ$} & \colhead{$^\circ$} & \colhead{$^\circ$} & \colhead{$\rm km~s^{-1}$} & \colhead{$\rm km~s^{-1}$} & \colhead{$\rm km~s^{-1}$} & \colhead{$\rm km~s^{-1}$} &\colhead{kpc}& \colhead{kpc} & \colhead{kpc} & \colhead{kpc}& \colhead{kpc} & \colhead{mas/yr} & \colhead{mas/yr} 
}
\startdata
1&126.58&-0.22&224.47&0.5&20.88&0.5&108&259&5&11&-14.86&-6.72&3.66&10.4&1.2&7.0&-6.4\\
2&131.92&11.17&215.93&0.2&30.83&0.2&69&188&  &  &-12.51&-3.27&3.32& 6.5&0.6&8.6&-7.4\\
3&138.65&22.29&206.03&0.2&40.89&0.2&  &   &  &  &-12.75&-2.31&4.57&7.0&0.4&10.0&-7.3\\
4&144.25&30.09&197.00&0.2&47.54&0.2&-7&39&1&3.9&-12.83&-1.47&5.52&7.5&0.3&10.9&-6.5\\
5&157.92&44.19&172.30&0.2&57.24&0.2&-71&-88&2&5.3&-12.30&0.58&6.74&8.0&0.5&11.8&-2.4\\
6&163.69&48.32&161.95&0.2&59.02&0.2&-87&-124&2&9.2&-12.30&1.41&7.54&8.8&0.8&11.5&-0.6\\
7&217.50&57.51&99.95&1.0&55.00&1.0&     &     &  &  &-8.98&5.6&8.08&9.9&1.2&4.1&5.2\\
\enddata
\end{deluxetable} 
}

\clearpage

\begin{deluxetable}{rrrrrrrrrrrrrrrr}
\tabletypesize{\scriptsize}
\tablecolumns{16}
\footnotesize
\rotate
\tablecaption{Spectra of GD-1 Stream Candidates}
\tablewidth{0pt}
\tablehead{
\colhead{SpecID} & \colhead{N} & \colhead{$\alpha$} & \colhead{$\delta$} & \colhead{$v_{\rm gsr}$} & \colhead{$\sigma(v)$} & \colhead{RV} & \colhead{$g_0$} & \colhead{$(u-g)_0$} & \colhead{$(g-r)_0$} &\colhead{[Fe/H]} & \colhead{$\sigma(\rm [Fe/H])$} & \colhead{log g} & \colhead {$\sigma(\rm log g)$} &\colhead{$\mu_{l}$}  &\colhead{$\mu_{b}$}  \\
&  & \colhead{$^\circ$} & \colhead{$^\circ$} & \colhead{$\rm km~s^{-1}$} & \colhead{$\rm km~s^{-1}$} & \colhead{$\rm km~s^{-1}$} & \colhead{mag} & \colhead{mag} & \colhead{mag} &\colhead{dex} & \colhead{dex} & \colhead{dex} & \colhead {dex}  &\colhead{mas/yr}  &\colhead{mas/yr}}
\startdata
1154-53083-266&1&125.683891&-0.439561&113.4&7.4&264.7&19.107&0.909&0.275&-1.89&0.08&3.46&0.25&3.5&-4.3\\
1154-53083-145&1&126.577150&-0.439386&97.8&7.8&249.4&19.102&0.904&0.180&-2.14&0.07&3.62&0.23&12.9&-10.5\\
1154-53083-155&1&126.645407&-0.340477&109.0&3.6&260.3&18.117&0.935&0.275&-2.06&0.04&3.30&0.22&11.6&-3.9\\
1760-53086-339&2&131.406984&9.770413&65.8&14.5&186.9&18.389&0.909&0.227&-2.18&0.11&4.05&0.29&9.4&-8.4\\
2671-54141-432&2&132.157853&11.135851&77.3&6.0&193.9&18.780&0.841&0.406&-2.24&0.04&3.54&0.22&10.0&-5.2\\
2667-54142-427&2&132.367656&11.325495&70.5&4.0&186.5&17.433&0.987&0.420&-2.17&0.02&2.77&0.24&6.7&-7.4\\
2667-54142-604&2&133.073392&12.266884&62.3&2.5&175.1&15.859&1.159&0.486&-2.08&0.03&2.22&0.15&8.4&-6.1\\
2319-53763-347&3&138.399875&22.413022&38.1&11.6&114.5&19.451&0.843&0.243&-2.24&0.22&3.80&0.09&13.2&-7.7\\
2319-53763-358&3&138.450751&22.434572&40.4&4.7&116.7&17.996&0.865&0.259&-1.93&0.02&3.90&0.10&12.5&-9.7\\
2319-53763-387&3&138.669378&22.559736&48.9&11.8&124.7&19.603&0.874&0.385&-2.07&0.07&3.68&0.30&12.1&-7.9\\
2889-54530-311&4&142.787815&29.461085&-0.9&4.2&47.8&17.454&0.972&0.417&-1.98&0.04&3.23&0.28&10.7&-9.4\\
2914-54533-297&4&142.978804&29.136307&-3.0&3.9&46.9&18.139&0.925&0.189&-2.09&0.04&3.94&0.16&8.8&-9.3\\
2889-54530-240&4&143.339524&29.620398&-6.2&4.7&41.7&17.933&0.883&0.256&-2.14&0.03&3.63&0.17&9.6&-6.7\\
2889-54530-204&4&143.424098&29.036453&-3.6&5.1&46.5&17.822&0.872&0.271&-2.21&0.05&3.28&0.20&6.7&-4.5\\
2889-54530-215&4&143.453188&29.120632&-3.4&1.7&46.4&14.981&1.194&-0.237&-2.09&0.11&3.20&0.17&9.5&-10.9\\
2889-54530-238&4&143.552948&29.477508&-4.3&4.6&44.1&17.843&0.888&0.293&-2.21&0.01&3.60&0.18&8.4&-9.1\\
2889-54530-225&4&143.597836&29.802697&-6.1&1.7&41.0&14.845&1.196&-0.102&-1.98&0.06&3.34&0.09&13.7&-11.1\\
2914-54533-171&4&143.718962&29.874100&-3.6&8.0&43.2&19.432&0.747&0.285&-2.26&0.08&3.33&0.56&15.4&-8.7\\
2914-54533-458&4&144.107345&30.737924&-9.8&4.8&33.6&18.716&0.779&0.216&-2.26&0.05&4.41&0.10&13.1&-6.5\\
2914-54533-453&4&144.120941&30.947724&-1.4&4.4&41.2&17.955&0.827&0.220&-2.10&0.02&3.79&0.28&14.3&-6.3\\
2914-54533-509&4&144.275720&30.280468&-4.4&3.9&40.6&18.292&0.909&0.187&-2.25&0.01&3.46&0.31&9.5&-12.2\\
2914-54533-515&4&144.487234&30.410401&-4.7&3.6&39.8&18.247&0.855&0.232&-1.82&0.04&3.72&0.15&10.5&-1.6\\
2914-54533-540&4&144.561728&30.968385&-1.9&11.0&40.4&20.190&0.854&0.344&-1.95&0.10&3.46&0.21&12.6&-3.8\\
2567-54179-396&5&157.181542&44.514349&-67.9&6.8&-85.0&18.745&0.841&0.254&-2.08&0.04&3.78&0.19&13.9&1.8\\
2567-54179-246&5&157.565788&43.643316&-73.0&6.0&-87.1&18.630&0.934&0.167&-2.23&0.01&3.71&0.18&12.9&-0.9\\
2567-54179-238&5&157.691771&43.797748&-71.9&12.7&-86.7&19.546&0.908&0.206&-2.21&0.02&3.27&0.16&10.1&-5.6\\
2567-54179-189&5&157.817344&43.923322&-70.1&11.9&-85.5&19.761&0.772&0.342&-1.88&0.07&4.23&0.36&9.0&3.5\\
2567-54179-211&5&157.889540&43.500286&-65.2&10.0&-79.0&19.309&0.901&0.205&-1.89&0.15&3.74&0.07&9.2&-0.3\\
2567-54179-212&5&158.051293&43.644022&-67.6&5.2&-82.1&18.349&0.839&0.208&-1.90&0.07&3.67&0.26&9.7&-0.2\\
2567-54179-170&5&158.137638&44.086320&-58.2&9.8&-74.4&19.112&0.863&0.201&-2.04&0.06&3.64&0.17&14.4&-1.7\\
2567-54179-491&5&158.437074&44.466389&-69.2&9.1&-87.1&18.242&0.880&0.209&-2.02&0.09&3.95&0.16&8.9&-1.9\\
2567-54179-511&5&158.700621&44.530392&-68.9&4.4&-87.2&18.083&0.954&0.237&-2.22&0.01&3.18&0.25&12.1&1.3\\
2557-54178-498&5&158.747249&44.686776&-74.2&3.9&-93.2&16.794&1.048&0.476&-2.12&0.02&2.60&0.27&13.5&2.7\\
2410-54087-297&6&162.371533&47.207633&-87.0&6.9&-118.3&18.893&0.885&0.255&-2.26&0.02&3.46&0.20&15.4&-1.3\\
2410-54087-236&6&162.451620&48.002695&-85.2&4.2&-119.5&18.305&0.896&0.250&-1.71&0.04&3.51&0.21&8.6&4.8\\
2410-54087-288&6&162.547222&46.990942&-88.0&12.2&-118.7&19.344&0.893&0.164&-2.08&0.08&3.78&0.14&9.6&-3.8\\
2390-54094-256&6&162.557763&47.251097&-95.0&2.6&-126.6&16.945&1.048&0.036&-1.83&0.00&3.89&0.19&16.1&-1.5\\
2410-54087-380&6&162.743760&48.903984&-85.3&5.1&-123.1&18.632&0.840&0.269&-1.80&0.07&3.81&0.05&12.7&-1.6\\
2410-54087-173&6&163.799206&47.813588&-88.8&7.1&-123.6&18.828&0.845&0.200&-2.06&0.07&3.86&0.30&8.9&-1.8\\
2410-54087-539&6&164.064054&49.034036&-92.2&8.2&-131.6&19.293&0.925&0.169&-1.68&0.11&3.10&0.30&8.1&-1.7\\
2410-54087-501&6&164.334475&48.472850&-81.1&7.8&-118.8&19.401&0.925&0.202&-1.76&0.08&3.80&0.28&7.3&0.0\\
2410-54087-504&6&164.371325&48.224481&-90.6&8.8&-127.4&19.409&0.841&0.208&-1.75&0.07&4.30&0.14&11.3&-0.6\\
2390-54094-565&6&164.616981&49.202693&-81.1&2.0&-121.6&16.919&1.105&0.487&-1.86&0.01&2.93&0.19&6.9&3.7\\
2390-54094-615&6&165.199453&48.669262&-92.4&3.9&-131.5&17.895&0.896&0.385&-2.24&0.02&3.81&0.16&11.8&-1.7\\
2410-54087-637&6&165.381730&48.671045&-87.1&6.2&-126.4&18.928&0.887&0.206&-2.29&0.05&3.84&0.12&7.8&-0.2\\
2539-53918-406&7&217.430060&59.067783&-163.8&3.0&-297.5&15.805&1.226&-0.133&-2.11&0.03&3.29&0.12&5.9&3.7\\
2539-53918-070&7&219.681591&57.897316&-148.9&10.7&-283.7&17.970&0.862&0.199&-2.15&0.02&3.73&0.04&2.5&10.8\\
2539-53918-596&7&219.898491&58.258368&-157.6&5.5&-293.3&17.035&1.197&-0.167&-2.10&0.26&2.92&0.43&1.0&2.9\\
\enddata
\end{deluxetable}

\clearpage

\begin{deluxetable}{cc}
\tabletypesize{\scriptsize} \tablecolumns{2} \footnotesize
\tablecaption{Fixed Galactic Potential Parameter Values} \tablewidth{0pt}
\tablehead{
\colhead{Parameter} & \colhead{Value} \\ 
}
\startdata
$\alpha$ & $1.0$ \\
$M_{disk}$ & $1.0 \times 10^{11} M_{\sun}$ \\
$M_{bulge}$ & $3.4 \times 10^{10} M_{\sun}$ \\
$a$ & $6.5$ kpc \\
$b$ & $0.26$ kpc \\
$c$ & $0.7$ kpc \\
$v_{halo}$ & $114$ km/s \\
$q$ & 1.0 \\
$d$ & 12 kpc \\
\enddata
\label{parameters}
\end{deluxetable}

\clearpage

{
\begin{deluxetable}{cccc}
\tabletypesize{\scriptsize} \tablecolumns{2} \footnotesize
\tablecaption{Fit Parameters} \tablewidth{0pt}
\tablehead{
\colhead{Fit Parameter} & \colhead{Description} & \colhead{Step Size} & \colhead{Random Selection Range}\\
}
\startdata
$R_5$ & Sun-centered distance of Region 5& 0.1 kpc & 5 : 13 kpc\\
$v_{x,5}$ & X Velocity of Region 5& 1 km/s & -130 : -60 km/s\\
$v_{y,5}$ & Y Velocity of Region 5& 1 km/s & -320 : -230 km/s\\
$v_{z,5}$ & Z Velocity of Region 5& 1 km/s & -80 : -120 km/s\\
\enddata
\label{fitpars}
\end{deluxetable}
}
\clearpage

\clearpage

\setcounter{page}{1}

\plotone{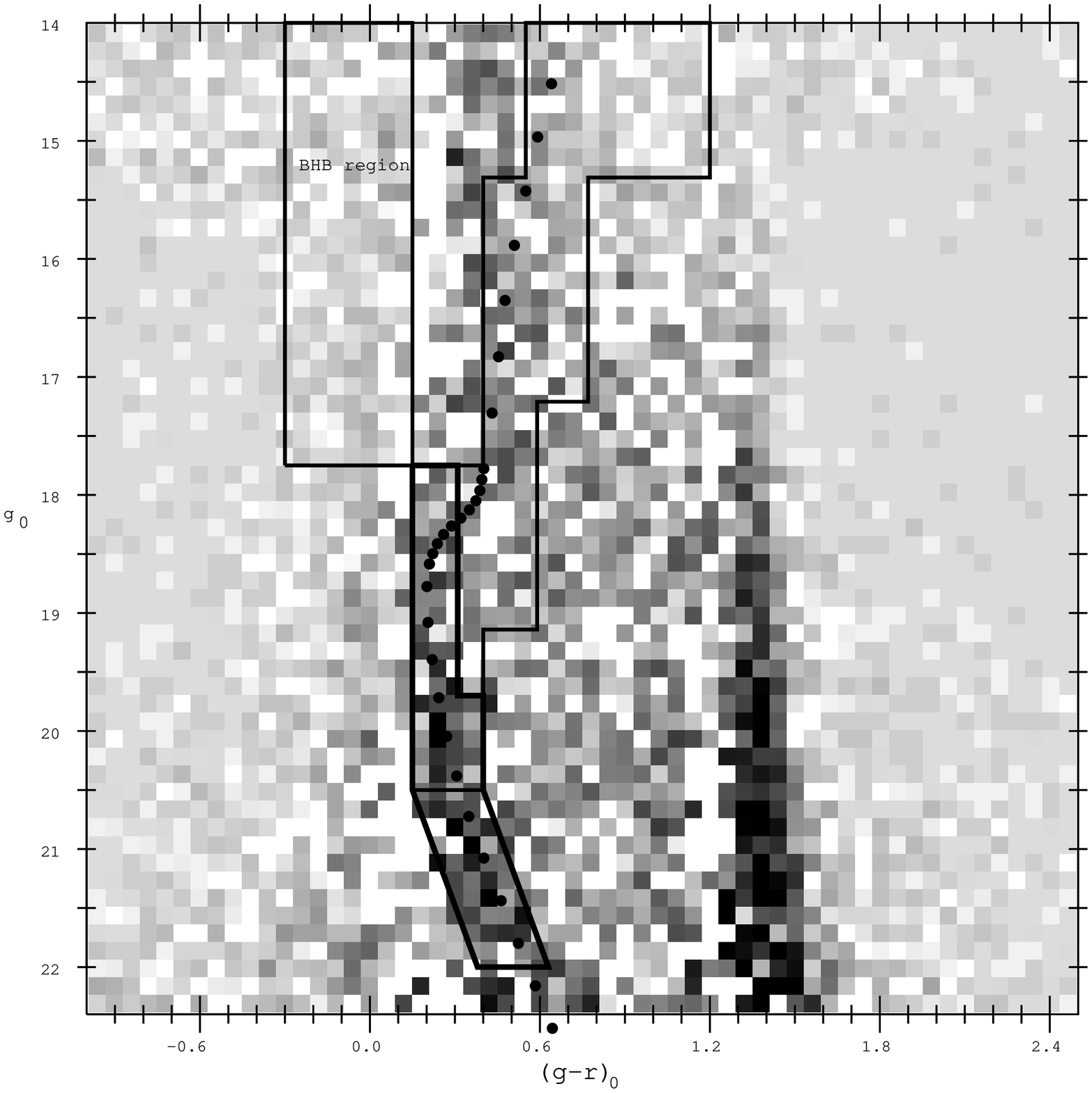}
\plotone{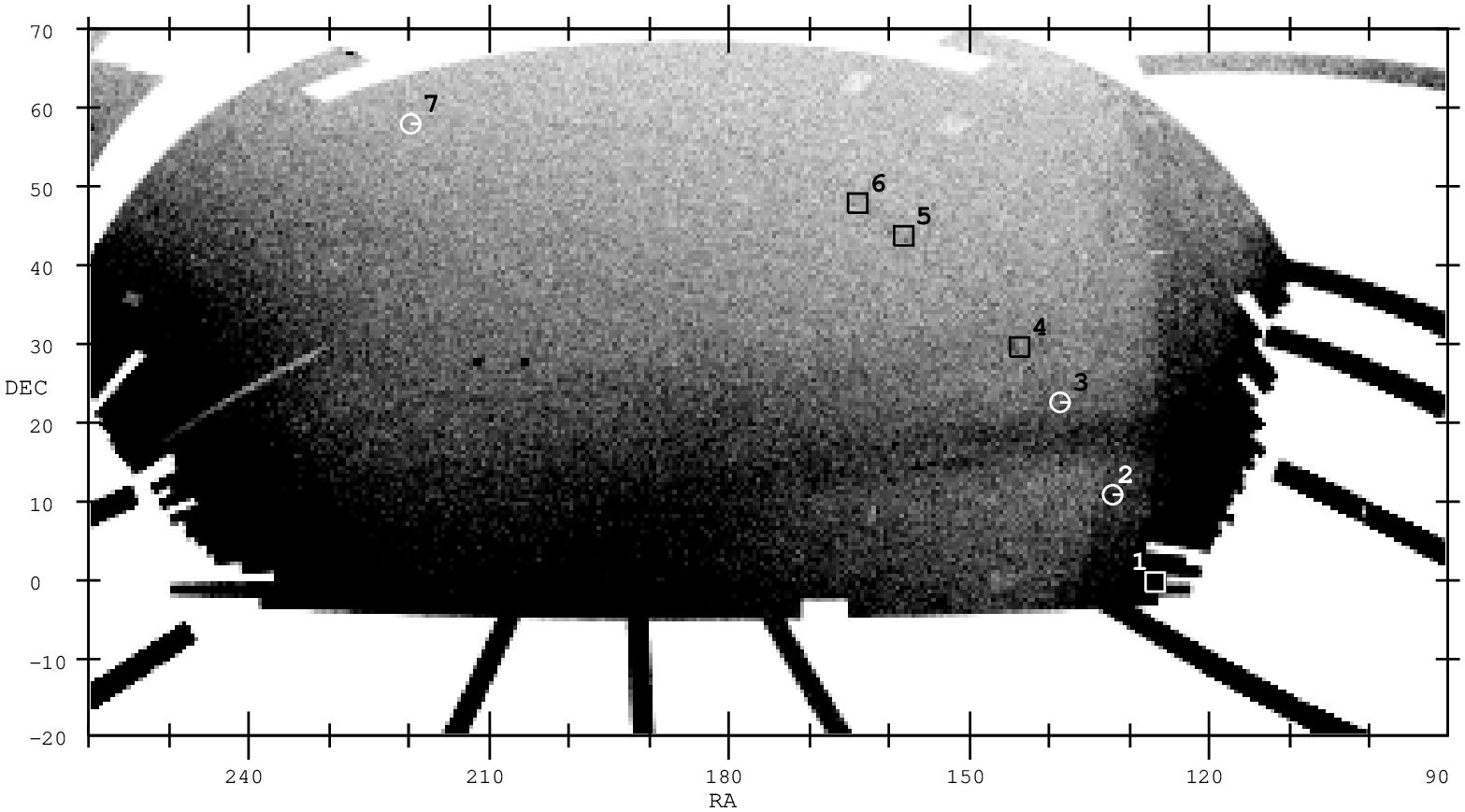}
\plotone{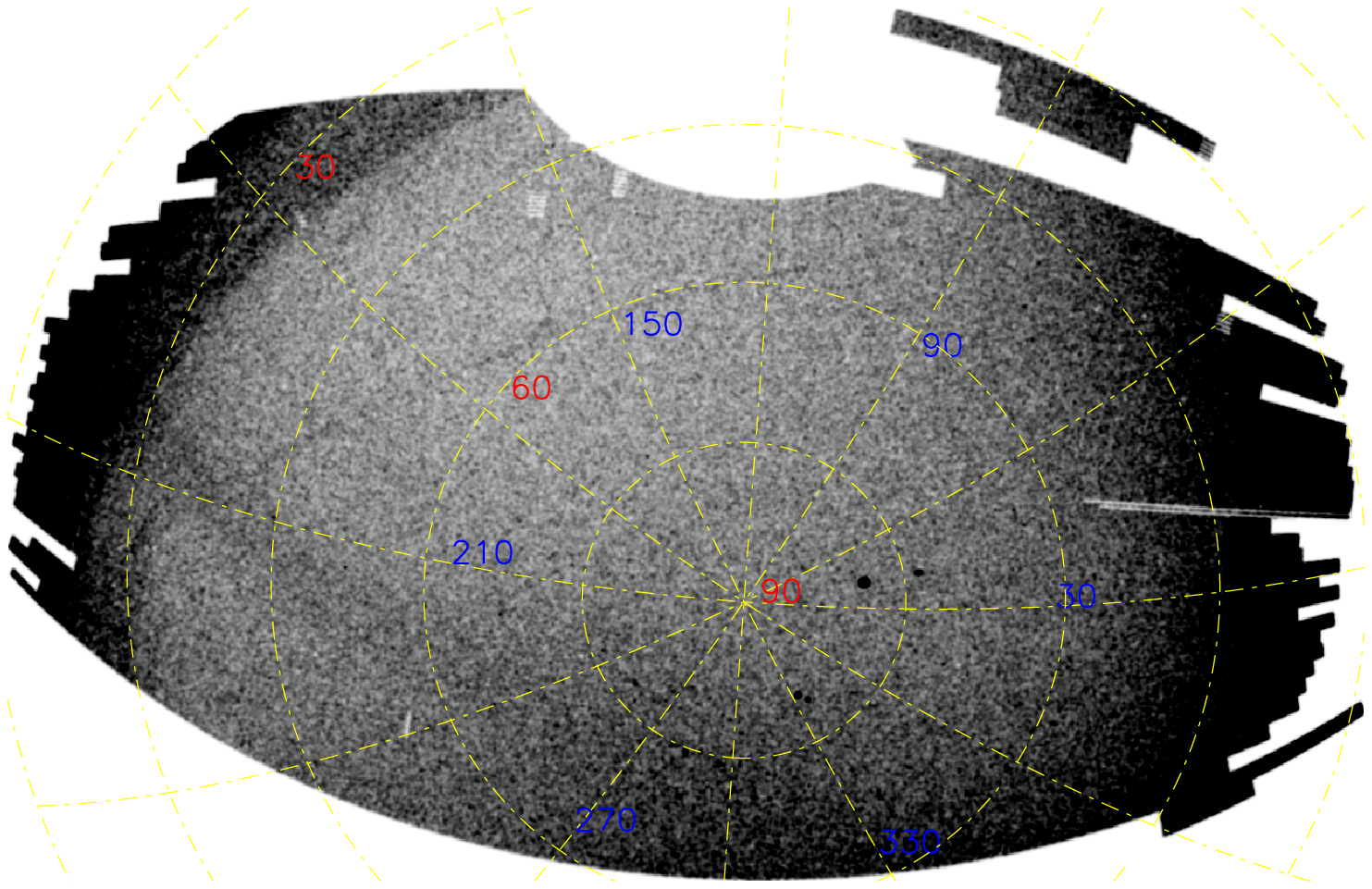}
\plotone{fig3.ps}
\plotone{fig4.ps}
\plotone{fig5.ps}
\plotone{fig6.ps}

\plotone{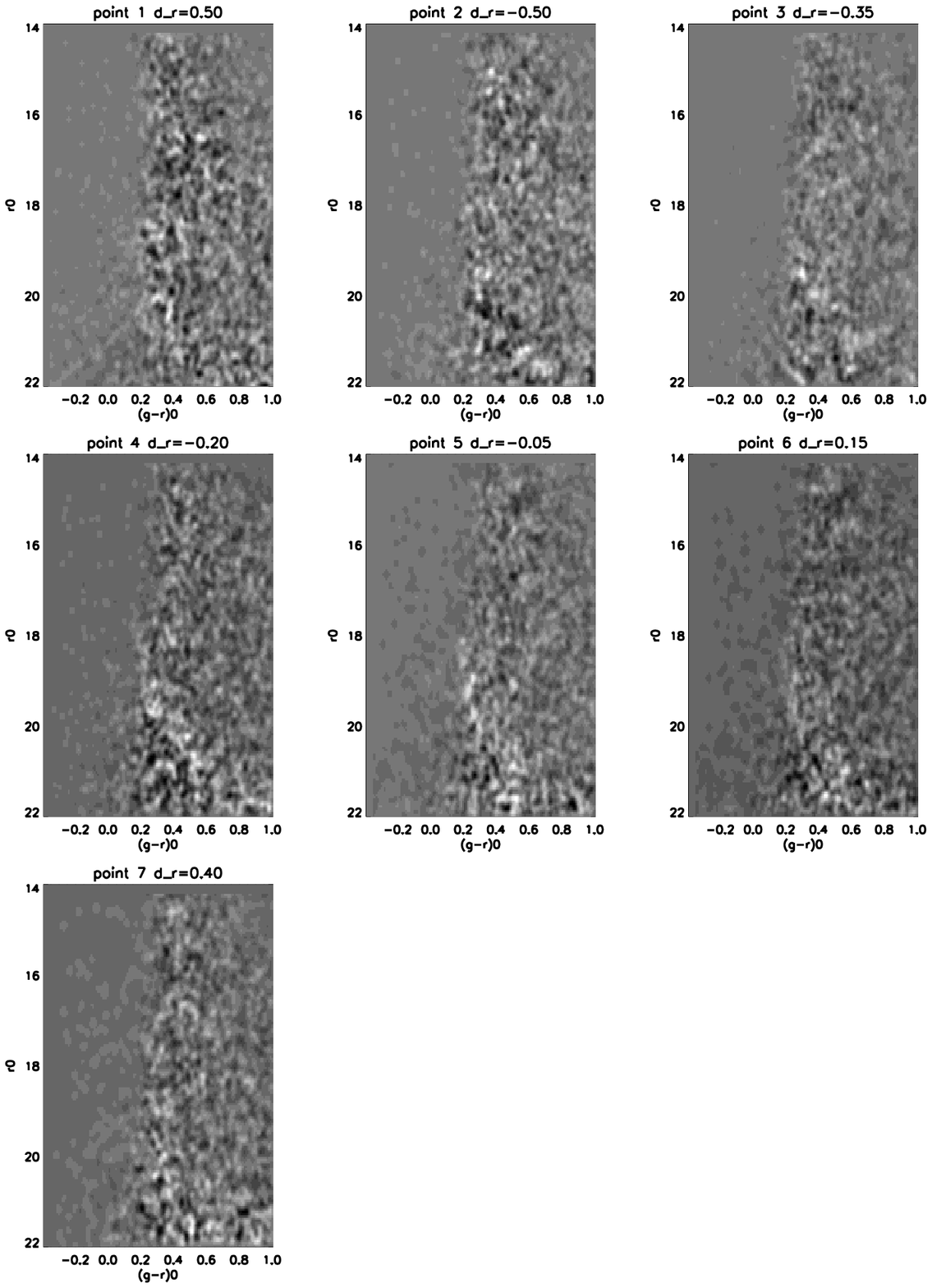}

\plotone{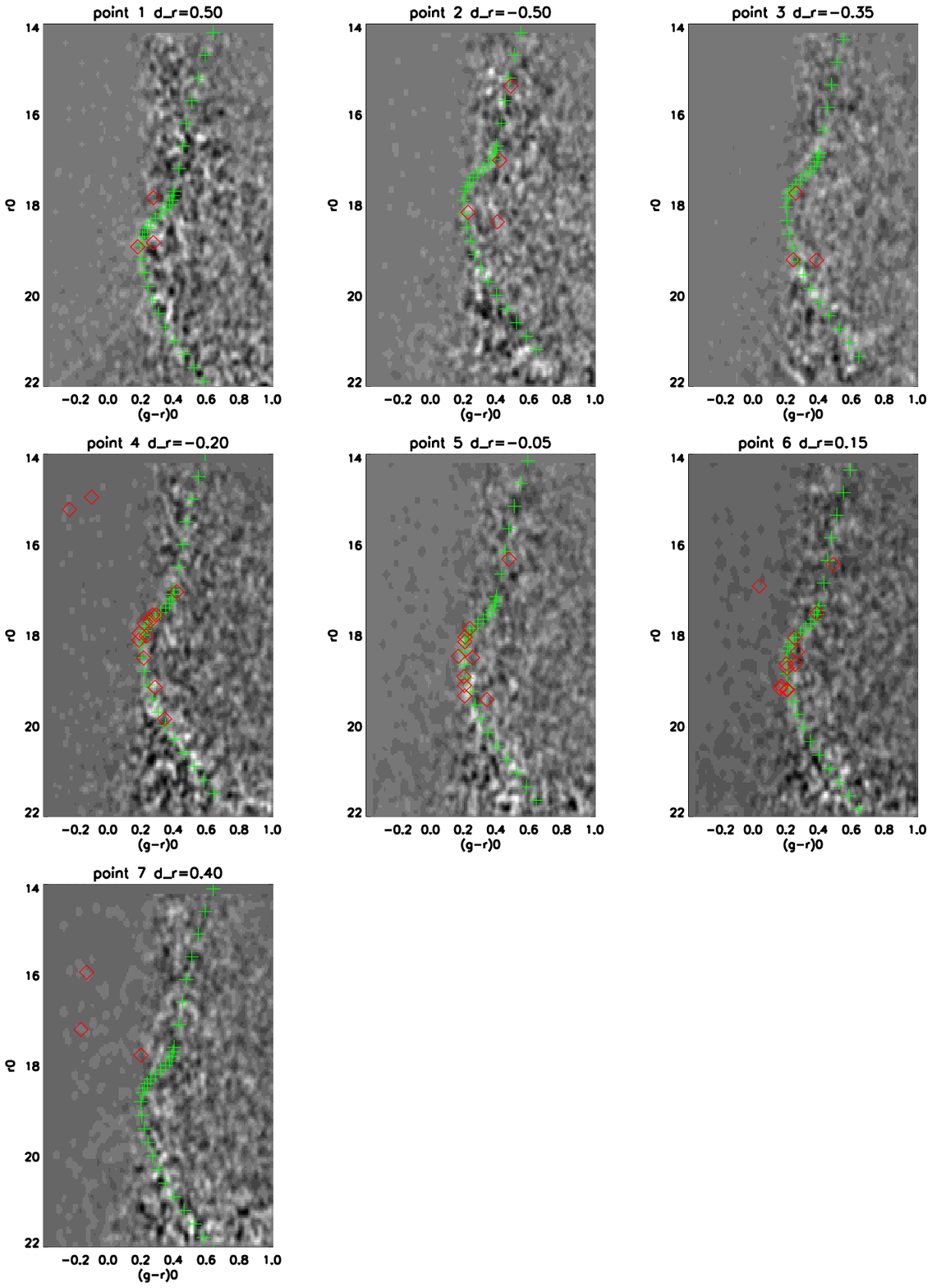}

\plotone{fig9.ps}
\plotone{fig10.ps}

\plotone{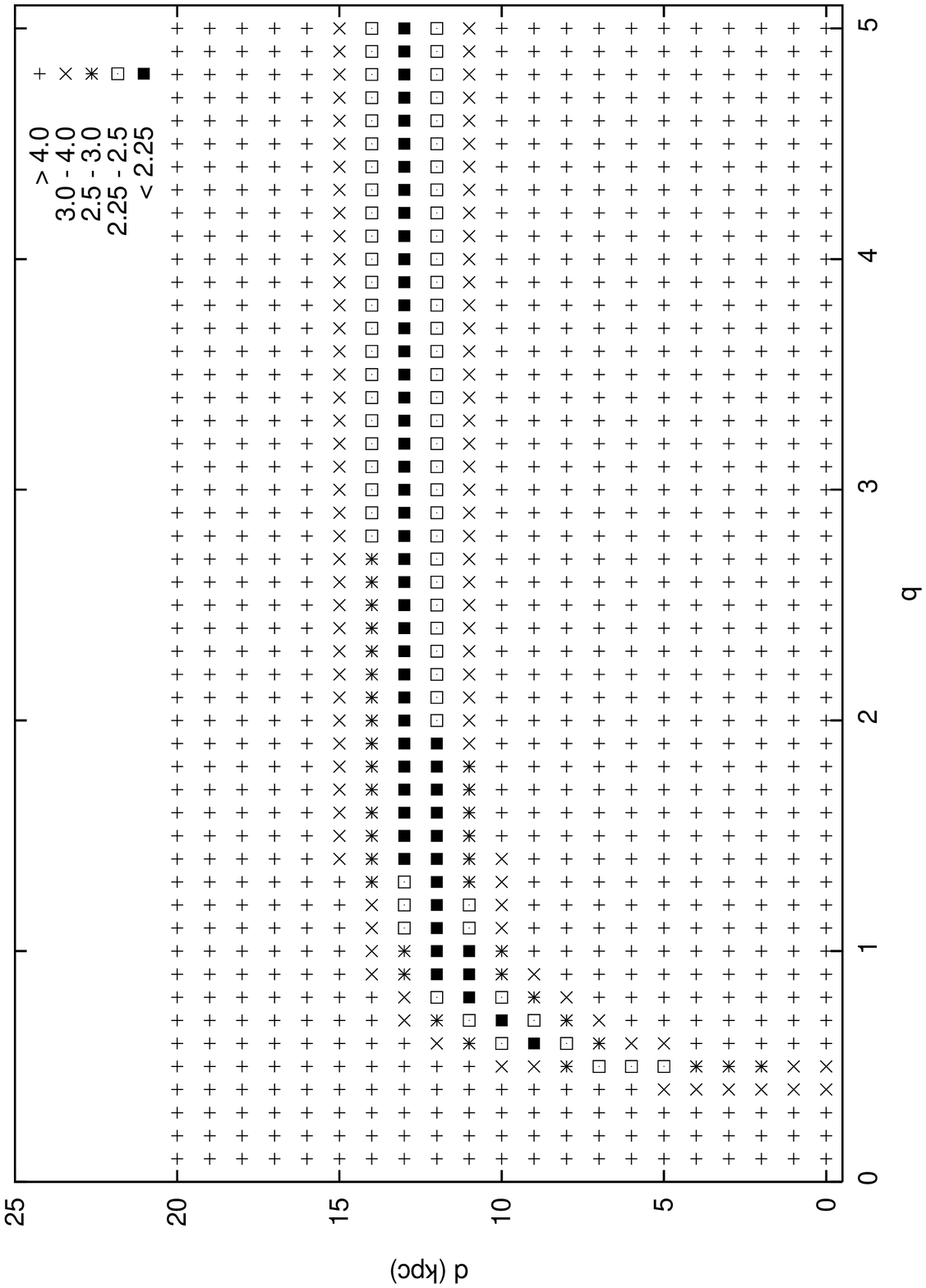}
\plotone{fig12.ps}
\plotone{fig13.ps}
\plotone{fig14.ps}

\end{document}